\documentclass[11pt,sort&compress]{elsarticle}

\usepackage{amssymb}
\usepackage{amsmath}
\usepackage{graphicx}
\usepackage{tabularx}
\usepackage{cite}
\usepackage{color}
\usepackage{fullpage}
\usepackage{subfigure}
\usepackage{subfigmat}
\usepackage{etoolbox}
\usepackage{url}
\usepackage{float}
\usepackage{bm}
\usepackage[dvipsnames]{xcolor}
\usepackage{placeins}%FloatBarrier 
\usepackage{siunitx} % for the units
\usepackage[nameinlink,capitalize]{cleveref}
\newcommand{\new}[1]{{\color{black}#1}}

\journal{Physics of Fluids}
\begin{document}
% % %------------------------------------------------------------------------------
\begin{frontmatter}

  \title{Numerical investigation of non-condensable gas effect on vapor bubble collapse}

  \author{Theresa Trummler$^{1,2}$}\ead{theresa.trummler@unibw.de}
  \author{Steffen J. Schmidt$^{1}$}
  \author{Nikolaus A. Adams$^{1}$}

  \address{
      $^{1}$Chair of Aerodynamics and Fluid Mechanics, Technical University of Munich \\ 
      Boltzmannstr.\ 15, 85748 Garching bei M\"unchen, Germany\\ 
      $^{2}$Institute of Applied Mathematics and Scientific Computing, Bundeswehr University Munich\\ Werner-Heisenberg-Weg 39, 85577 Neubiberg, Germany }

  \begin{abstract}
      We numerically investigate the effect of non-condensable gas inside a vapor bubble on bubble dynamics, collapse pressure and pressure impact of spherical and aspherical bubble collapses. Free gas inside a vapor bubble has a damping effect that can weaken the pressure wave and enhance the bubble rebound. To estimate this effect numerically, we derive and validate a multi-component model for vapor bubbles containing gas. For the cavitating liquid and the non-condensable gas, we employ a homogeneous mixture model with a coupled equation of state for all components. The cavitation model for the cavitating liquid is a barotropic thermodynamic equilibrium model. Compressibility of all phases is considered in order to capture the shock wave of the bubble collapse. After validating the model with an analytical energy partitioning model, simulations of collapsing wall-attached bubbles with different stand-off distances are performed. The effect of the non-condensable gas on rebound and damping of the emitted shock wave is well captured.
  \end{abstract}

\end{frontmatter}

 \noindent\fbox{%
    \parbox{0.99\textwidth}{%
        The following article has been accepted by \textit{Physics of Fluids}. After it is published, it will be found at the journal's website.
    }%
}

\section{Introduction}
\label{s:Introduction} 

  In technical applications and experiments, it can be assumed that a certain amount of non-condensable gas is present in vapor cavities. In general, gases are dissolved in liquids~\citep{Pollack:1991ue} and are released during pressure reduction by outgassing~\citep{Iben:2015bw,Freudigmann:2017cr} or cavitation~\citep{Franc:2004fu}. In experiments with cavitation bubbles, gases are produced when the bubbles are generated with lasers or sparks through chemical reactions and recombination processes~\citep{Sato:2013fg,Akhatov:2001hy}.

  Gas inside a vapor bubble has a damping effect that can weaken the pressure wave and increase the rebound of the bubble. For spherical bubble collapses, the damping effect is evident in the incompressible Rayleigh-Plesset equation~\citep{plesset1949dynamics} 
  \begin{equation}
    \rho_l (\ddot{R}R+ 3/2\,\dot{R}^2)= - \Delta p + p_{g}, 
      \label{eq:rp}
  \end{equation}
  here written in inviscid form neglecting surface tension, with the density of the liquid $\rho_l$, the bubble radius $R$, its derivate $\dot{R}$, the driving pressure difference $\Delta p=p_{\infty}-p_{sat}$, and the gas pressure $p_{g}=p_{g,0}\left(R_0/R\right)^{3\gamma}$. $p_{g,0}$, $R_0$, $\gamma$ denote the initial gas content, the initial bubble radius and the adiabatic index, respectively. The compressible Keller-Miksis Equation~\citep{Keller1980:wm} additionally captures the rebound. Taking advantage of the fact that it can be treated first order~\citep{Prosperetti:1987jt} and neglecting viscosity and surface tension, it simplifies to 
  \begin{equation}
    \rho_l (\ddot{R}R (1-v)+ 3/2 \dot{R}^2(1-v/3))=\nonumber \\
    (- \Delta p+p_{g})(1+v)+R\dot{p_{g}}/c_l,
  \label{eq:km}
  \end{equation}
  with $v=\dot{R}/c_l$. $c_l$ is the speed of sound in the liquid phase. Both equations clearly show that the partial pressure of the gas inside the bubble decelerates the collapse and, in the compressible formulation, enhances the rebound. Further, both effects are more pronounced at lower driving pressure differences $\Delta p$.

  Analytical studies evaluating the effect of gas inside vapor bubbles were conducted by \citet{Fujikawa:1980jj} and \citet{Akhatov:2001hy}. They studied bubble dynamics of vapor bubbles containing gas and considered compressibility, non-equilibrium effects at phase transition, and conductive heat transfer. Later, \citet{Tinguely:2012wo} experimentally investigated the effect of the driving pressure on the energy partitioning into shock wave energy and rebound energy for spherical bubble collapses under microgravity. Based on their findings, they derived an analytical model from the Keller-Miksis Equation predicting the energy partitioning based on one single non-dimensional parameter, which also depends on the gas content in the bubble. While the effect of gas and driving pressure on the collapse of spherical bubbles has already been investigated analytically and experimentally, for more complex configurations, however, the effect has not yet been elucidated. In experimental studies, it is challenging to determine or control the initial gas content in the bubble. Additionally, \new{the short time scale and the high intensity of the emitted pressure wave impose high requirements on the measurement equipment~\citep{Tinguely:2012wo}, and more accurate measurements have only recently become feasible \citep{Supponen:2017wl,supponen2019high,supponen2019detailed} }. Three-dimensional\new{,} time-resolved numerical simulations\new{, in which the gas content can be precisely controlled and the pressure signals monitored,} are \new{thus well suited for complementary and detailed studies of the effect of gas in complex configurations.} 

  In the last decade, \new{compressible} numerical simulations have become a complementary tool for studying collapse dynamics\new{\citep{Johnsen:2009cua, Lauer:2012jh}}. \new{Several numerical studies~\citep{Johnsen:2009cua, Beig:2018ga,Pishchalnikov:2018pp,Trummler:2020JFM} focused on the effect of the first collapse and considered gas bubbles neglecting phase transition. \citet{Pishchalnikov:2018pp} varied the gas content in an elliptical, wall-attached bubble and showed how this affects the collapse behavior and the pressure impact.} To capture \new{both} the pressure waves emitted at collapse and \new{the} rebound, the modeling approach must account for both compressibility and phase transition. Previous works on bubble collapses considering both employed an equilibrium cavitation model in combination with a single-fluid approach as e.g. \citet{Sezal:2009diss, ochiai2011numerical, Pohl:2015keb, Oerley:2016diss,Koukouvinis:2016ir} \new{and more recently \citet{sagar2020dynamics,Trummler:2021if}}. For vapor bubbles containing gas, a multi-component model considering a cavitating liquid and an additional gas component is necessary. \citet{Orley:2015kt} extended the barotropic equilibrium cavitation model by \citet{Schnerr:2008jja} and \citet{Schmidt:2009} by an additional non-condensable gas component. In this model, the mass fraction of gas is convected and for all components a coupled equation of state is employed. So far, the multi-component model has been applied and validated for the injection of a cavitating liquid into a gaseous ambient. Many research groups have taken up the model and partly modified it. \citet{Orley:2016db} extended the model to employ different equations of state for the individual components; \citet{mithun2018numerical} added a volume-of-fluid method for interface capturing; \citet{brandao2020numerical} considered a finite-rate mass transfer for the cavitation process\new{.} 

  In this work, we present an adaptation of the multi-component model of \citet{Orley:2015kt} and \citet{Trummler:2018AAS} to be applicable to vapor bubbles containing gas. Preliminary studies to this work were presented in \citet{Trummler:2018ww,Trummler:2019icmf}. \new{In this paper, we first introduce the thermodynamic model and then apply it to spherical and aspherical bubble collapses.}
  For the simulations of the aspherical collapses, we have chosen a driving pressure of 1 bar. As \cref{eq:rp,eq:km} show, $\Delta p$ governs the intensity of the emitted pressure wave, the rebound, and the influence of the gas. Under atmospheric conditions, a stronger rebound and a more pronounced damping effect of the gas occur than, for example, at 100 bar. Further, the choice is also motivated by the fact that experiments of single bubble collapses are often conducted at atmospheric conditions, see e.g. \citet{Philipp:1998eg,dular2019high}, and we can thus ensure better comparability. 

  An important parameter for aspherical collapses is the stand-off distance. The stand-off distance has a significant influence on the collapse dynamics and the erosion potential as has been shown by experimental\new{\citep{Tomita:1986gy,Philipp:1998eg}} and numerical studies~\new{\citep{Lauer:2012jh,Trummler:2020JFM,Trummler:2021if}}. The sign of the stand-off distance alters the collapse behavior and a smaller stand-off distance (absolute value) increases the pressure impact on the wall. Therefore, we consider wall-attached bubbles with negative and positive stand-off distances. 

  The paper is organized as follows. In \cref{s:Methods}, we describe the physical model and the numerical method. \Cref{s:spherical} presents simulation results of spherical bubble collapses with various gas contents and driving pressures, and the validation of the modeling approach with the analytical energy partitioning model by \citet{Tinguely:2012wo}. Then, in \cref{s:aspherical}, we present and analyze simulation results of collapsing wall-attached bubbles at different stand-off distances with and without gas. \Cref{s:ConclusionAndDiscussion} summarizes the paper.

\section{Physical Model and Numerical Method}
\label{s:Methods}

  \subsection{Governing Equations}
  \label{subsec:GovEq}

  We solve the fully compressible Navier-Stokes equations and an additional transport equation for the gas mass fraction 
      \begin{equation}
          \partial_{t}\boldsymbol{U}+\nabla \cdot [ \boldsymbol{C}(\boldsymbol{U})+\boldsymbol{S}(\boldsymbol{U})]=0\,. 
          \label{eq:NS}
      \end{equation}
  The state vector $\boldsymbol{U}=[\rho , \, \rho \boldsymbol{u},\,\rho\xi_g ]^T$ is composed of the conserved variables density $\rho$ and momentum $\rho \boldsymbol{u}$ and gas density $\rho\xi_g$. Due to the assumed barotropic modeling ($p=p(\rho)$), the energy equation can be omitted. The convective fluxes $\boldsymbol{C}(\boldsymbol{U})$ and the flux contributions due to pressure and shear $\boldsymbol{S}(\boldsymbol{U})$ read
  \begin{equation}
     \boldsymbol{C}(\boldsymbol{U})=
              \boldsymbol{u}
                 \begin{bmatrix} 
                    \rho\\ 
                    \rho\boldsymbol{u}\\ 
                    \rho\xi_g
              \end{bmatrix}
              \quad
              \mathrm{and}
              \quad
              \boldsymbol{S}(\boldsymbol{U})=
              \begin{bmatrix} 
              0\\ 
               p \boldsymbol{I}-\boldsymbol{\tau}\\ 
               0\\
              \end{bmatrix},
                   \label{eq:NS_Basis}
      \end{equation}    
  with the velocity $\boldsymbol{u}$, the static pressure $p$, the unit tensor $\boldsymbol{I}$, and the viscous stress tensor $\boldsymbol{\tau}$
  \begin{equation}
    \boldsymbol{\tau}=\mu(\nabla \boldsymbol{u}+(\nabla \boldsymbol{u})^{T}-\frac{2}{3}(\nabla \cdot\boldsymbol{u})\boldsymbol{I}),
    \label{eq:tau}
  \end{equation}
  where $\mu$ is the dynamic viscosity.

  \subsection{Thermodynamic Model}
  \label{subsec:ThermoModel}
  
  We adopt a multi-component homogeneous mixture model~\citep{Orley:2015kt,Trummler:2018AAS} to be applicable to vapor bubbles containing gas. In the employed modeling approach, the cavitating liquid ($lv$) and the non-condensable gas ($g$) are described by a substitute mixture fluid. This approach implies that within a computational cell all phases have the same velocity, temperature and pressure. The single fluid is described by the volume averaged density inside a computational cell 
  \begin{equation}
  \rho= \sum \beta_{\phi} \rho_{\phi}=\beta_{g}\rho_{g}+(1-\beta_{g})\rho_{lv}.
  \label{eq:rhomix}
  \end{equation}
  $\beta_{\phi}$ denotes the volume fraction and $\rho_{\phi}$ the density of each component \new{$\phi= \{ lv,\, g\}$}. The gas volume fraction $\beta_{g}$ can be obtained from the transported mass fraction $\xi_{g}$ by the following relation 
  \begin{equation}
   \beta_{g}=\xi_{g}\frac{\rho}{\rho_{g}}. 
  \label{eq:beta_g}
  \end{equation}
  For the mixture fluid a coupled equation of state (EOS) is derived. Therefore corresponding thermodynamic relations for each component are derived. 

  For the modeling of vapor bubbles containing gas, the pressure acting on the liquid-vapor mixture in the bubble has to be modified. Inside the bubble the pressure is composed of the partial pressures of vapor and gas as 
  \begin{equation}
   p = p_{lv} + p_{g}.
  \end{equation}
  We calculate the pressure acting on the liquid-vapor mixture by 
  \begin{equation}
  p_{lv} = p - p_{g} = (1-\beta_{g})\,p.
  \end{equation}

The cavitating water is described with a barotropic EOS, derived by integration of the isentropic speed of sound
\begin{equation}
 \rho_{lv}=\rho_{\mathrm{sat},l}+(p_{lv}-p_\mathrm{sat}) / c^2, 
 \label{eq:rho_lv}
\end{equation}
where \new{$\rho_{\mathrm{sat},l}$} is the saturation density for liquid water and $p_\mathrm{sat}$ the saturation pressure. Phase change is modeled assuming local thermodynamic equilibrium. For $p_{lv}>p_\mathrm{sat}$, there is purely liquid water and $c=1482.35\,\si{m/s}$. For $p_{lv}<p_\mathrm{sat}$, there is a liquid vapor mixture with $c=0.1\,\si{m/s}$ as a typical value for an equilibrium isentrope, see e.g. \citet{Franc:2004fu}. The vapor volume fraction $\alpha$ is given by the density of the liquid-vapor mixture \new{$\rho_{lv}$} as 
    \begin{equation}
        \alpha=\frac{\rho_{\mathrm{sat},l}-\rho_{lv}}{\rho_{\mathrm{sat},l}-\rho_{\mathrm{sat},v}}. 
    \end{equation}
\new{Note that $l$ refers to liquid and $v$ to vapor. }       
For water at reference temperature $\mathrm{T}=\SI{293.15}{K}$, the corresponding values are $p_\mathrm{sat}=\SI{2340}{Pa}$, $\rho_{\mathrm{sat},l}=\SI{998.1618}{kg/m^3}$ and $\rho_{\mathrm{sat},v}= 17.2\cdot 10^{-3}\,\si{kg/m^3}$. 

The non-condensable gas phase is described with 
\begin{equation}
\rho_{g}=\rho_{g,\mathrm{ref}}(p/p_\mathrm{ref})^{1/\gamma},
\label{eq:rho_g}
\end{equation}
where $\rho_{g,ref}$ is the reference density at the reference pressure $p_\mathrm{ref}$. Here we used $p_\mathrm{ref}=10^5\,\si{Pa}$ and $\rho_{g,\mathrm{ref}}=1.188\,\si{kg/m^3}$. In the results presented, the gas is modeled as isothermal with $\gamma=1$.

By inserting the thermodynamic relations for each component (\cref{eq:rho_lv}, \cref{eq:rho_g}) in \cref{eq:rhomix} a coupled EOS $p=p(\rho, \xi_g)$ is derived, see \citet{Orley:2015kt}. 

Viscous effects are considered in our simulations using a linear blending of the volume fractions for the mixture viscosity. The following values for the viscosities are used: $\mu_{l} = 1.002\cdot 10^{-3}\,\si{\pascal\second}$ , $\mu_{v} = 9.272\cdot 10^{-6}\,\si{\pascal\second}$ and $\mu_{g} = 1.837\cdot 10^{-5}\,\si{\pascal\second}$\,. 

\subsection{Numerical Method}
The thermodynamic model is embedded in a density-based fully compressible flow solver with a low-Mach-number-consistent flux function, see \citet{Schmidt:2015wa}. For the reconstruction at the cell faces an upwind biased scheme is used, where the velocity components are reconstructed with the up to third-order-accurate limiter of \citet{Koren:1993} and the thermodynamic quantities $\rho$, $p$ with the second-order minmod slope limiter of~\citet{Roe:1986}.

Time integration is performed with an explicit second-order, 4-step low-storage Runge-Kutta method~\citep{Schmidt:2015wa}. 

\section{Spherical collapses and validation of the modeling approach}
\label{s:spherical}

To validate the modeling approach, we simulate spherical collapses of vapor bubbles containing various amounts of gas. We analyze the collapse and rebound behavior and the intensity of the emitted pressure wave. In \cref{ss:val} the model is compared with the energy partitioning model of \citet{Tinguely:2012wo}.

\subsection{Set-up}

\new{We consider a bubble with an initial radius $R_{0}=400\,\si{\micro\metre}$. Note that previous investigations have shown that the normalized rebound~\citep{Akhatov:2001hy} and the energy partitioning~\citep{Tinguely:2012wo} are independent of the bubble size. The bubble} is placed at the center of a box with dimension $500 \times R_{0}$ in each Cartesian direction. Taking advantage of symmetry, only an eighth of a bubble is simulated. The domain is discretized with an equidistant grid within a cubic sub-domain with an edge length of $1.25\; R_{0}$, and for the outer part a grid stretching is applied. Simulations are performed on different grid levels defined by the number of cells over the initial radius $N_C/R_{0}$. If not stated otherwise, the results are for a grid-resolution of $N_C/R_{0}=80$. The pressure field is initialized with a pressure jump at the pseudo phase boundary. A constant CFL number of 1.4 is used. 

\begin{figure}[!tb]
 \centering
   \subfigure[]{\includegraphics[height=4cm]{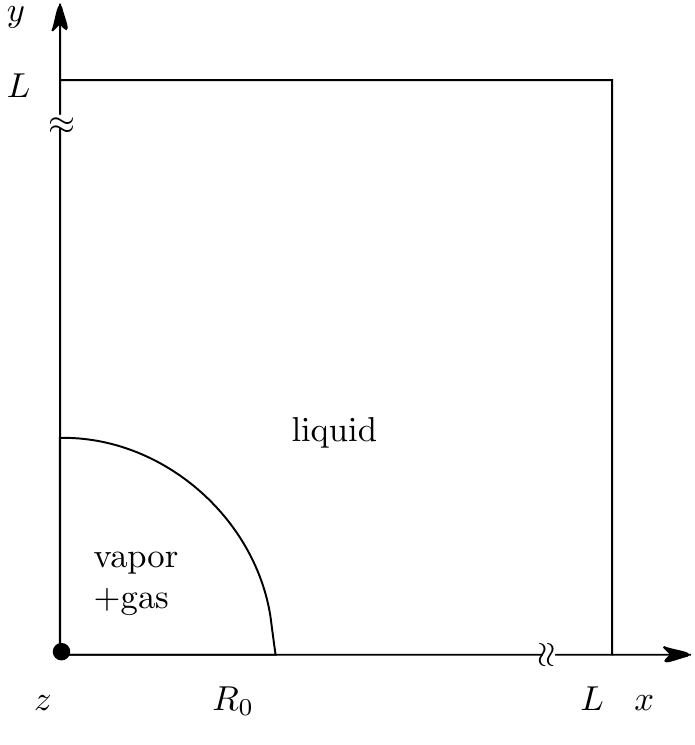}}
   \subfigure[]{\includegraphics[height=4cm]{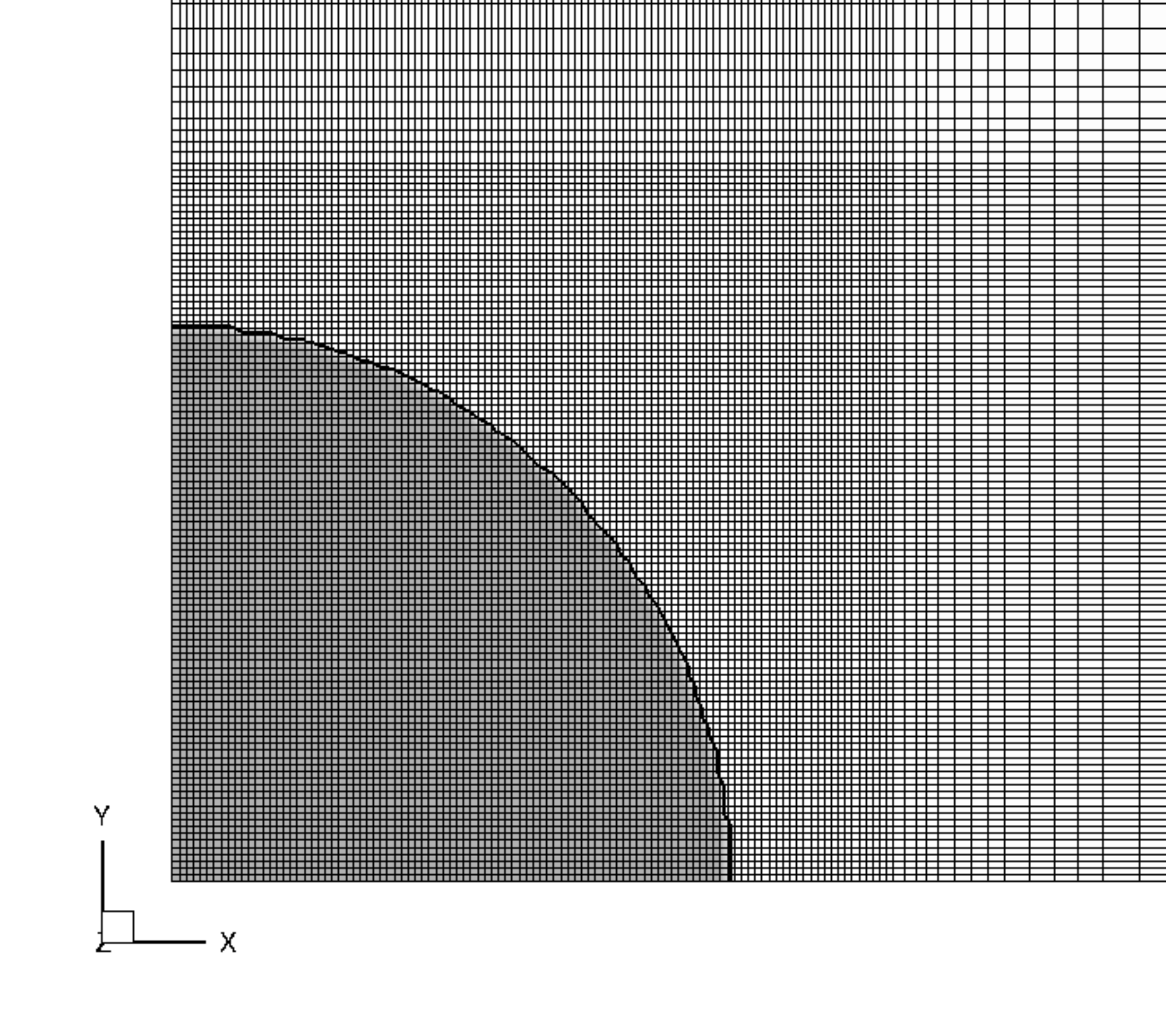}}
 \caption{Simulation set-up. (a) Planar sketch of the numerical set-up, (b) Grid in near bubble region and initialized bubble.}
 \label{fig:setup_sph}
\end{figure}

For this investigation, the initial gas content in the bubble $p_{g,0}$ and the driving pressure difference $\Delta p$ are varied covering different combinations of $\Delta p = [10^4\,\si{Pa},\, 10^5 \,\si{Pa}]$ and $p_{g,0} =[0\,\si{Pa},\;1000 \,\si{Pa}]$. During the simulations, pressure signals are recorded at certain radial positions from the bubble center. 

\new{In this section (\cref{s:spherical}), time is normalized with the Rayleigh collapse time for spherical collapses~\citep{Rayleigh:1917}
\begin{equation}
        \tau_c=0.915\cdot R_{0}\sqrt{\rho_l / \Delta p}.  
        \label{eq:t_c}
\end{equation}
Following previous studies~\citep{Beig:2018ga,Trummler:2020JFM,Trummler:2021if}, the pressure is normalized in both sections (\cref{s:spherical,s:aspherical}) using 
\begin{equation}  
    p^{\ast} = c_l \sqrt{\rho_l \Delta p}. 
\end{equation} }

\subsection{Results}
\label{ss:results_spherical}
\begin{figure}[!tb]
 \centering
   \subfigure[]{\includegraphics[width=0.7\linewidth]{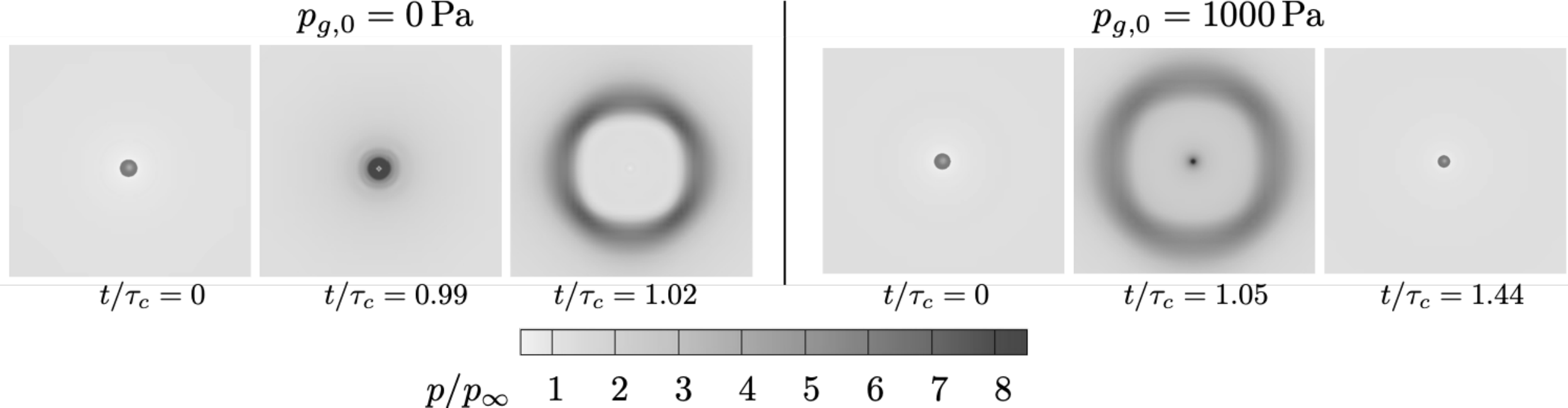}}\\
   \subfigure[]{\includegraphics[width=0.7\linewidth]{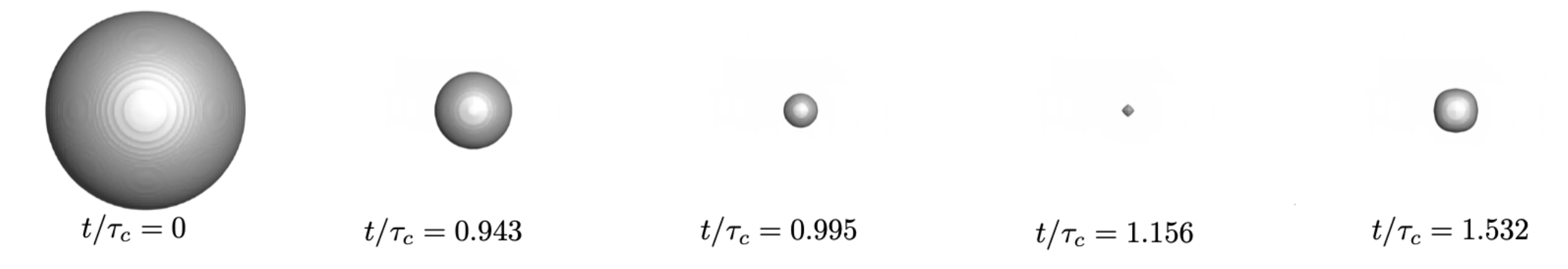}}
 \caption{Time series of bubble collapse and rebound. (a) Pressure field for 
 $\Delta p = 10\,\si{kPa}$ with $p_{g,0} = 0\,\si{Pa}$ (left) and with $p_{g,0} = 1000\,\si{Pa}$ (right), (b) Near bubble region to visualize the rebound for $\Delta p = 10 \,\si{kPa}$ with $p_{g,0} = 1000 \,\si{Pa}$.}
 \label{fig:co_dyn}
\end{figure}

\Cref{fig:co_dyn}~(a) depicts the bubble collapse and the rebound at different time steps for $\Delta p = 10\,\si{kPa}$ with $p_{g,0} = \{ 0 \;\si{Pa}, \,1000 \;\si{Pa}\}$. The left time series presents the bubble collapse without gas, showing the initial bubble, the situation shortly before the collapse and the emitted shock wave after collapse. Analogously, the dynamics of a bubble with a high gas content is visualized in the right time series. In this case, a rebound is visible at $t = 1.44\;\tau_{c}$. In \cref{fig:co_dyn}~(b) the near bubble region is shown to visualize the rebound behavior. \new{As can be seen in the last two time instants ($t/\tau_c=1.156$ and $t/\tau_c=1.532$), the rebound bubble is not completely spherical, which is due to a more accurate numerical reconstruction in the direction of the grid orientation.}

\begin{figure*}[!tb]
 \centering
   \includegraphics[height=7.0cm]{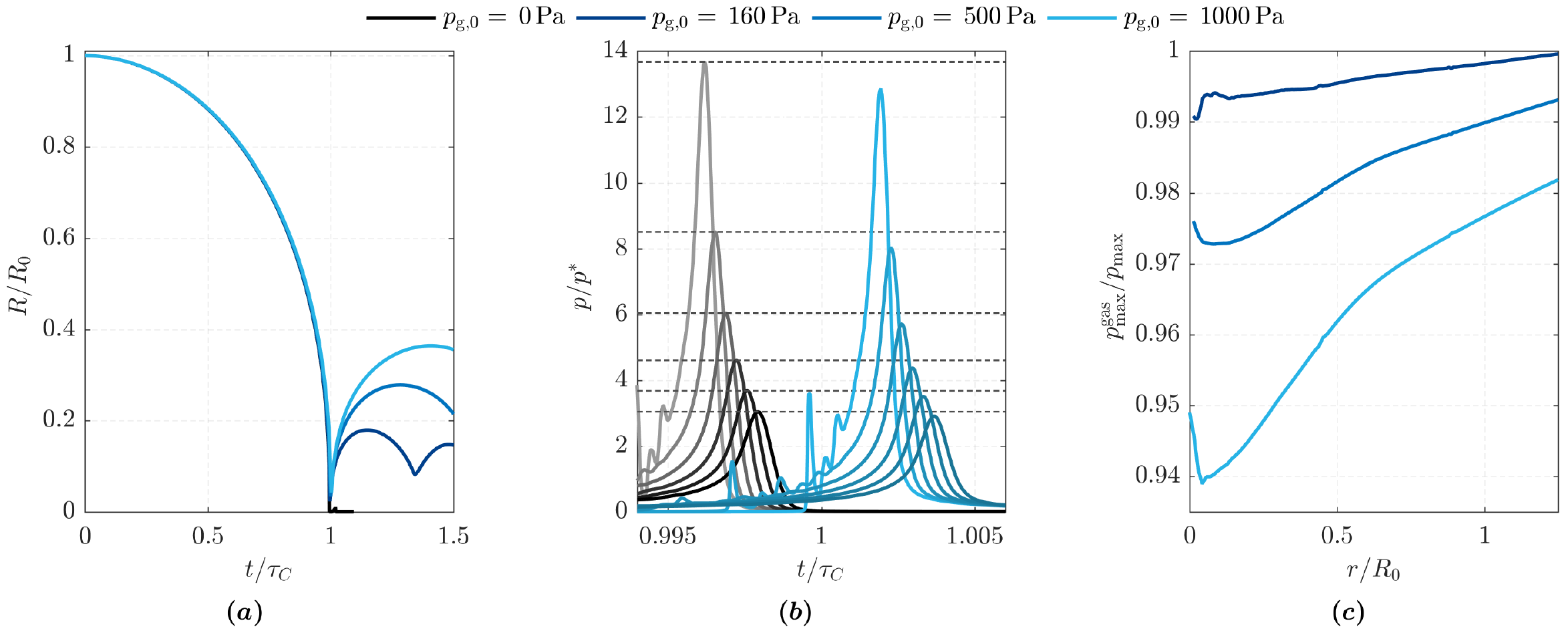}
 \caption{Simulation results. (a) Temporal evolution of the bubble radius \new{for different $p_{g,0}$ (see legend) ;} (b) Pressure signals from the probes $0.1:0.05:0.35\,R_0$ \new{with line color gradation corresponding to the probe position for the cases $p_{g,0} = 0\,\si{Pa}$ (gray ($0.1\,R_0$) to black ($0.35\,R_0$) ) and $p_{g,0} = 1000\,\si{Pa}$ (light blue ($0.1\,R_0$) to blue ($0.35\,R_0$));} (c) Maximum pressure compared to that without gas. (Grid resolution 80 $N_C/R_0$).}
 \label{fig:results}
\end{figure*}

\Cref{fig:results}~(a) compares the temporal evolution of the normalized bubble radius $R/R_{0}$ for different gas contents. In configurations with gas, the bubbles rebound significantly. Besides the rebound, the non-condensable gas in the vapor bubble also affects the intensity of the emitted pressure wave. \Cref{fig:results}~(b) shows the monitored pressure at certain radial positions from the bubble center and different gas contents. The radial decay of the maximum pressure is obvious and the presence of gas reduces the maximum pressure. The damping effect of the gas is more distinct for probes closer to the bubble center. Additionally, the pressure signals reveal that the collapse time is closely matched. \Cref{fig:results}~(c) compares the pressure maximum in the near bubble region. Again, the damping effect of the gas and the decay of the damping effect with increasing distance to the focus point are evident. 

\begin{figure*}[!tb]
 \centering
    \subfigure{\includegraphics[height=7.0cm]{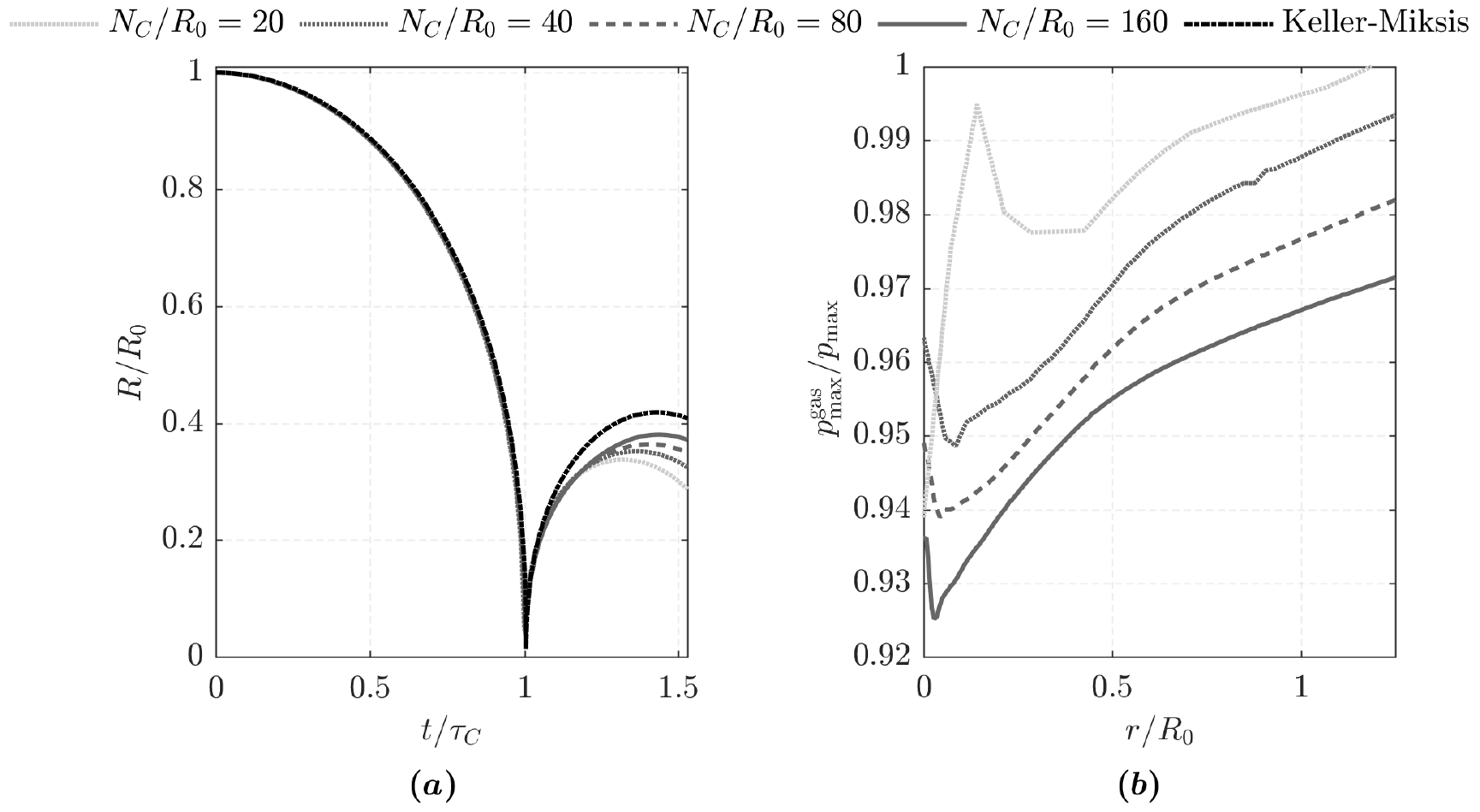}}
 \caption{Grid effect on the rebound and the damping of the maximum pressure by the gas for $\Delta p=10^5\,\si{Pa}$ and $p_g=1000\,\si{Pa}$. (a) Temporal evolution of the bubble radius, (b) Maximum pressure compared to that without gas. }
 \label{fig:grid}
\end{figure*}

The grid resolution is known to affect the minimum bubble radius and the rebound~\citep{Beig:2018ga,schmidmayer19, Trummler:2018ww} and the intensity of the pressure peaks~\citep{Mihatsch:2015db,Schmidt:2014ev,Trummler:2018ww}. To assess the grid influence, we have conducted a grid study. \Cref{fig:grid}~(a) depicts the temporal evolution of the bubble radius for different grid resolutions. As expected, the rebound increases with increasing grid resolution and approaches the one predicted by the Keller-Miksis equation. \Cref{fig:grid}~(b) compares the maximum pressure of the configuration with gas ($p_{max}^{gas}$) to that without gas ($p_{max}$). At all grid resolutions, the gas has a damping effect on the maximum pressure, although a higher grid resolution results in higher damping since the focus point is better resolved and the transport of the emitted shock wave is less dissipative. \new{In conclusion, both the rebound and the damping of the maximum wall pressure show a grid dependence, leading to a more pronounced gas effect on higher grid resolutions. However, as discussed and shown in \citet{Trummler:2018ww} and illustrated here in \cref{fig:grid}, the gas effect is already captured on the coarsest grid resolution of 20 $N_C/R_0$. On a grid resolution of 80 $N_C/R_0$ (\cref{fig:results}), the gas effect is clearly pronounced for the considered $p_{g,0}$.} Based on our observations, we consider a grid resolution of 80 $N_C/R_0$ as a good compromise between accuracy and computational cost. 

\subsection{Validation with Energy Partitioning Model}
\label{ss:val}

\citet{Tinguely:2012wo} experimentally and theoretically investigated the effects of the driving pressure difference $\Delta p$ and initial gas content $p_{g,0}$ on bubble dynamics and shock wave emission. They postulated that the initial energy of a bubble $E_{0}$ mainly partitions into rebound energy $E_{reb}$ and shock wave energy $E_{sw}$
 \begin{equation}
 E_0 = E_{reb} + E_{sw}, 
\label{eq:e0}
\end{equation}
which is in terms of normalized energies $\epsilon_{reb} =E_{reb}/E_0$, $\epsilon_{sw} =E_{sw}/E_0$ 
 \begin{equation}
   \epsilon_{reb}+\epsilon_{sw}=1.
\label{eq:eps}
\end{equation}
The initial energy and the rebound energy are potential energies at the corresponding time instants~\citep{obreschkow2006cavitation}
\begin{equation}
E_{0} = \frac{4 \pi }{3} R_0^3 \Delta p \quad \mathrm{and} \quad E_{reb} =\frac{4 \pi }{3} R_{reb}^3 \Delta p,
\label{eq:Ereb}
\end{equation}
and thus the normalized rebound energy $\epsilon_{reb}$ is 
 \begin{equation}
    \epsilon_{reb} =E_{reb}/E_0=\left(R_{reb}/R_{0}\right)^3 .
    \label{eq:eps_reb}
\end{equation} 
The shock wave energy $E_{sw}$ at a distance $d$ from the focus point reads~\citep{Vogel:b15KVKtx}
 \begin{equation}
    E_{sw} =\frac{4 \pi d^2}{\rho_l c_l} \int p(t)^2 dt .
    \label{eq:e_sw}
\end{equation} 
Based on the assumption that the pressure signals $p(t)$ have a universal shape that scales with the peak value $p_{max}$, one can estimate $E_{sw}\propto p_{max}^2$. \new{Hence, the normalized shock wave energy $\epsilon_{sw}$ can be 
assessed by the relative damping of the peak values as
 \begin{equation}
    \epsilon_{sw} \approx (p_{max}/p_{max,no\,rebound})^2.
    \label{eq:eps_sw2}
\end{equation} 
Alternatively, the normalized shock wave energy $\epsilon_{sw}$ can be approximated using \cref{eq:eps} with
 \begin{equation}
  \epsilon_{sw} \approx 1 - \epsilon_{reb}.
  \label{eq:eps_sw1}
 \end{equation}}

\begin{figure*}[!tb]
 \centering
  \includegraphics[width=0.65\linewidth]{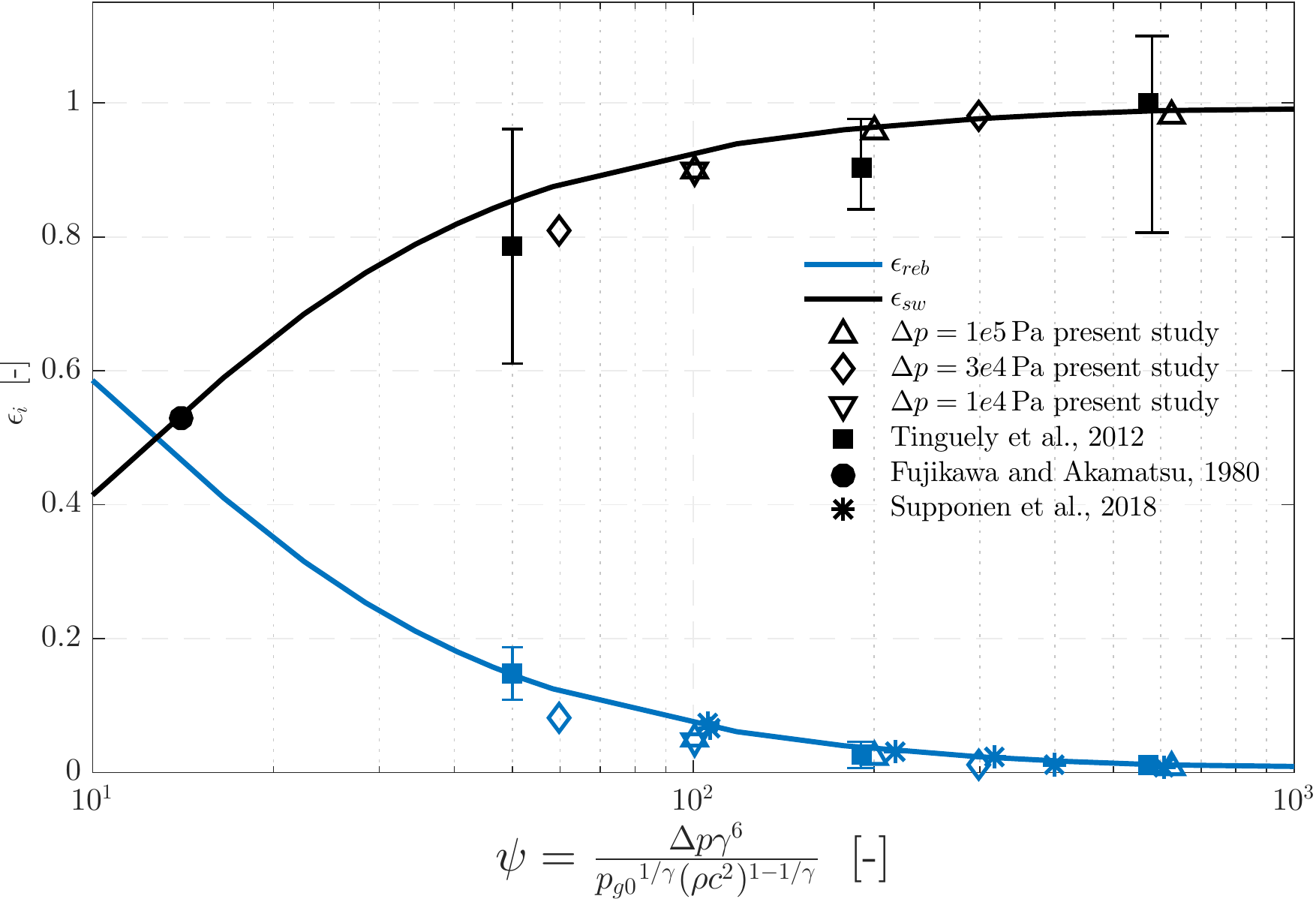}
  \caption{Simulation results in comparison with the theoretical energy partitioning proposed by \citet{Tinguely:2012wo} and \new{data from the literature. The solid curves are the results from the theoretical model. Filled symbols refer to data from the literature and empty ones to simulation results, where the color corresponds to the energy. The experimentally obtained values by \citet{Tinguely:2012wo} ({\tiny $\blacksquare $}) are shown along with the measurement error bars. The data of \citet{Supponen:2018tv} ($*$) consists only of rebound data and we have assumed $p_{g,0}=1.5\,\si{Pa}$ and $\gamma=1.4$. }}
 \label{fig:energies}
\end{figure*}

\citet{Tinguely:2012wo} derived a theoretical model using the inviscid Keller-Miksis equation (\cref{eq:km}) to predict the energy partitioning. 
Based on this model and experimental measurements, they were able to show that the energy fractions of rebound $\epsilon_{reb}$ and shock wave energy $\epsilon_{sw}$ depend on a single parameter
\begin{equation}
 \psi = \frac{\Delta p \gamma^{\,6}}{{p_{g,0}}^{1/\gamma}(\rho_l c_l^2)^{1-1/\gamma}}\, .
\label{eq:xi_energy}
\end{equation}
\Cref{fig:energies} plots the energy partitioning over $\psi$. The shock wave energy increases with $\psi$ and thus with the driving pressure difference and decreases with the partial pressure of free gas. On the other hand, the rebound is enhanced for a lower driving pressure difference and a higher gas content. 

\new{Additionally, experimental data of \citet{Tinguely:2012wo} including measurement error bars and data of \citet{Fujikawa:1980jj} are shown in \cref{fig:energies}. We also included bubble rebound data obtained for varying driving pressures by \citet{Supponen:2018tv}. They used partially degassed water and we have assumed $p_{g,0}=1.5\,\si{Pa}$ and $\gamma=1.4$. }

For the comparison of the simulation results with the energy partitioning model, the normalized rebound energy $\epsilon_{reb}$ is obtained from the maximum radius of the bubble in the first rebound using \cref{eq:eps_reb}. 
For the normalized shock wave energy $\epsilon_{sw}$, the pressure signals recorded at the bubble center are numerically integrated and set in relation to the respective values without gas and thus no rebound. Additionally, we have also evaluated the square of the ratios of the collapse pressures, see \cref{eq:eps_sw2}, and obtained comparable results. The evaluated energy partitioning from simulation data is included in \cref{fig:energies}. \new{At high $\psi$-values ($\psi \geq 200$), our simulation results agree very well with the theoretical model and literature data, while at lower $\psi$-values the simulation results show a smaller rebound and a higher damping effect than predicted by the theoretical model, see $\psi =60$. Thus, we conclude that our model is well suited to study configurations corresponding to high $\psi$-values with $\psi \geq 200$.} Further, the simulation data also show clear $\psi$-equivalence, i.e., equal $\psi$-values lead to equal normalized rebound and shock wave energies (see upward-pointing and downward-pointing triangles in \cref{fig:energies}). This successful validation allows for the application of the model to more complex configurations as the collapse of a wall-attached bubble.

\section{Aspherical collapse of a wall-attached bubble}
\label{s:aspherical}

\subsection{Set-up}

\begin{figure}[tb]
    \centering
        \includegraphics[width=0.4\columnwidth]{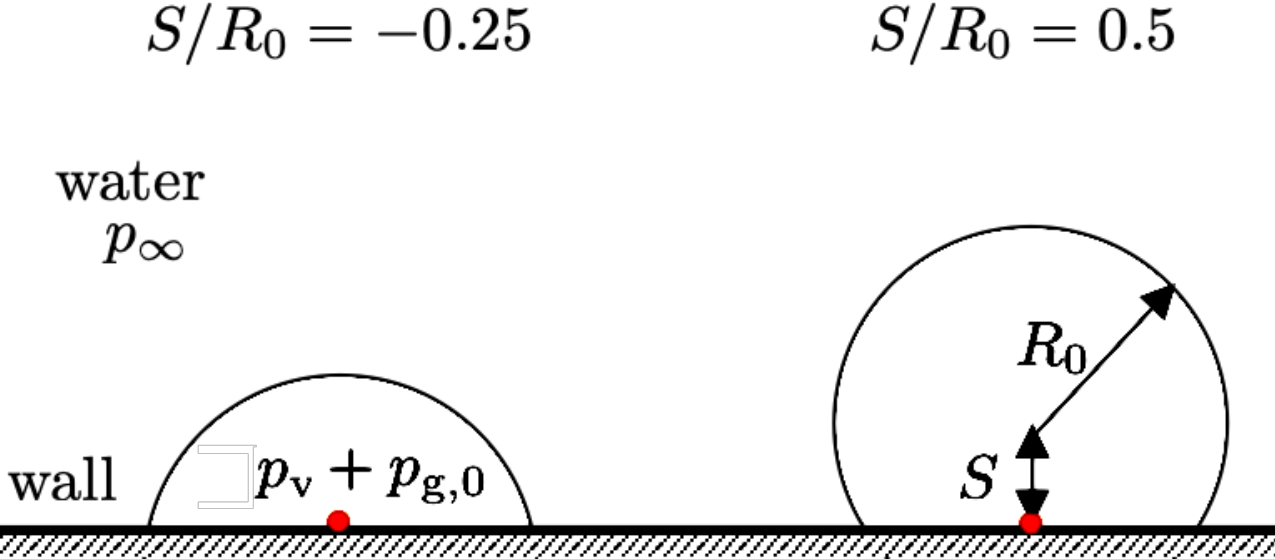}
    \caption{Sketch of the investigated configurations $S/R_0=-0.25$ and $S/R_0=0.5$. The red dot marks the position where the pressure signals are monitored.}
    \label{fig:setup_as}
\end{figure}

\new{\Cref{fig:setup_as} shows the investigated configurations with the two considered stand-off distances from the wall $S/R_0=-0.25$ and $S/R_0=0.5$. Following previous numerical studies~\citep{Lauer:2012jh,Oerley:2016diss,Koukouvinis:2016ir, Trummler:2021if,Trummler:2020JFM}, we consider an initial bubble radius $R_{0}$ of $400\,\si{\micro\metre}$. For non-spherical bubble collapses, it has been shown that the jet characteristics~\citep{Supponen:2016jnb} and the energy partitioning into rebound and shock wave energy~\citep{Supponen:2017wl,Supponen:2018tv} are determined by a dimensionless anisotropy parameter. In case of anisotropy due to nearby walls, this parameter is is independent of the bubble size and only a function of $S/R_0$.} We initialize the pressure field with a jump at the bubble interface with a driving pressure difference of $\Delta p =10^5\,\si{Pa}$. We consider either pure vapor bubbles ($p_{g,0} = 0 \,\si{Pa}$) or vapor bubbles containing a non-condensable gas content of $p_{g,0} = 160\,\si{Pa}$. This value is chosen based on the following considerations. Using experimental data, \citet{Tinguely:2012wo} estimated the initial partial gas content of non-condensable gas inside laser-generated bubbles in water to be $7\pm3.5\,\si{Pa}$. Since we model the gas as isothermal and not adiabatic, we decided to use the isothermal $ \psi$-equivalent of the lower limit of the estimated gas content of $p_{g,0}=3.5\,\si{Pa}$. For an adiabatic index of $\gamma=1.4$, a driving pressure of 1 bar and water, this value is $\psi = 630$. Thus, we consider the $\psi$ equivalent gas content for $\gamma=1$ which is $p_{g,0}=160\,\si{Pa}$.

Taking advantage of symmetry, only a quarter of the bubble is simulated. The bubble is placed in the center of a rectangular domain with an extension of $125\times R_{0}$ in wall-normal direction and $250\times R_{0}$ in wall-parallel directions. The domain is discretized with an equidistant grid within the near bubble region (80 $N_C$/$R_{0}$) and for the outer part a grid stretching is applied. The grid study presented in \cref{ss:results_spherical}
 demonstrated that this resolution is a good compromise of accuracy and required resources. In total, the grid has about 15 million cells. A constant CFL number of $1.4$ is used, which corresponds to a time step of $\Delta t \approx 1.5\,\si{\nano\second}$. 

To obtain dimensionless quantities, time is normalized with
\begin{equation}
        t^{\ast}= R_{0}\sqrt{\rho_l / \Delta p}\new{,}
        \label{eq:t_ast}
\end{equation}
\new{which is an estimate of the collapse time of a near-wall bubble collapse \citep{Plesset:1971hu}.}The wall has a retarding effect on the collapse and thus $t^{\ast}$ is longer than the Rayleigh collapse time for spherical collapses ($\tau_c=0.915\,t^{\ast}$\new{, see also \cref{eq:t_c}}). Velocity and pressure are normalized as
\begin{equation}  
    u^{\ast} = \sqrt{\frac{\Delta p}{\rho_l}}, 
    \quad \text{and} \quad
    p^{\ast} = c_l \sqrt{\rho_l \Delta p}.
\end{equation} 
\new{Note that $p^{\ast}$ corresponds to a water hammer pressure induced by the velocity $u^{\ast}$, see also \citet{Trummler:2021if}. The employed expression for $p^{\ast}$ can be related to the scaling found by \citet{Supponen:2017wl} for the maximum pressure at non-spherical bubble collapses. At a fixed stand-off distance (and thus anisotropy parameter), the maximum pressure measured at a distance $d$ from the focus point is 
\begin{equation}
  p_\mathrm{max}\propto c_l \sqrt{\rho_l \Delta p} (R_0/d)^{1.25}= p^{\ast}(R_0/d)^{1.25}.
  \label{eq:supponen_pmax} 
\end{equation} }

During the simulations, we monitor the integral vapor and gas volumes, the flow field at selected positions and evaluate the maximum pressure induced within the total simulation time. In the results, the pressure signals at the wall-center are presented. 

\subsection{Results}

In the following simulation results of a collapsing bubble with a negative stand-off distance (\cref{sss:neg}) and with a positive one (\cref{sss:pos}) are presented and discussed. 

\begin{figure*}
 \centering
    {\includegraphics[width=0.8\linewidth]{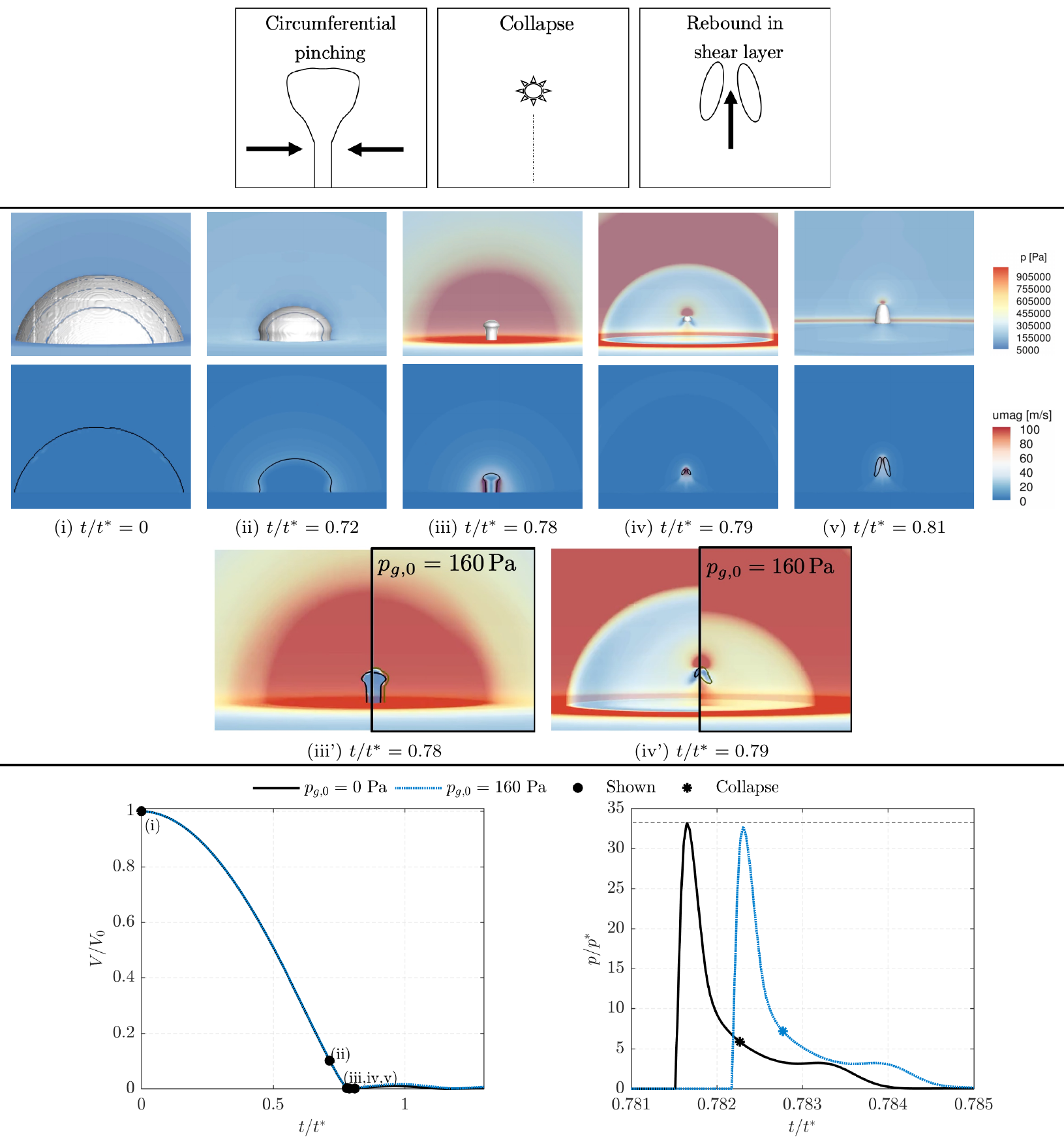}}
    \caption{Collapsing wall-attached bubble with $S/R_0=-0.25$. Top \new{panel}: Sketch of general collapse behavior. Middle \new{panel}: Time series showing pressure and velocity magnitude on midplane and isosurface/isoline 10\% vapor [(i)-(v)] and a comparison of $p_{g,0}=0\,\si{Pa}$ and $p_{g,0}=160\,\si{Pa}$ with additionally olive isolines 10\% gas [(iii)-(iv)]. Note that the discontinuities in the isosurface are due to post-processing issues. Bottom \new{panel}: Temporal evolution of bubble volume (left), and recorded pressure signals at the wall-center (right). \new{Reproduced from \citet{Trummler:2021diss}}. }
    \label{fig:m0100_ts}
\end{figure*}

\begin{figure*}
    \centering
    \subfigure[]{\includegraphics[trim={0 0 0 0},clip,height=6cm]{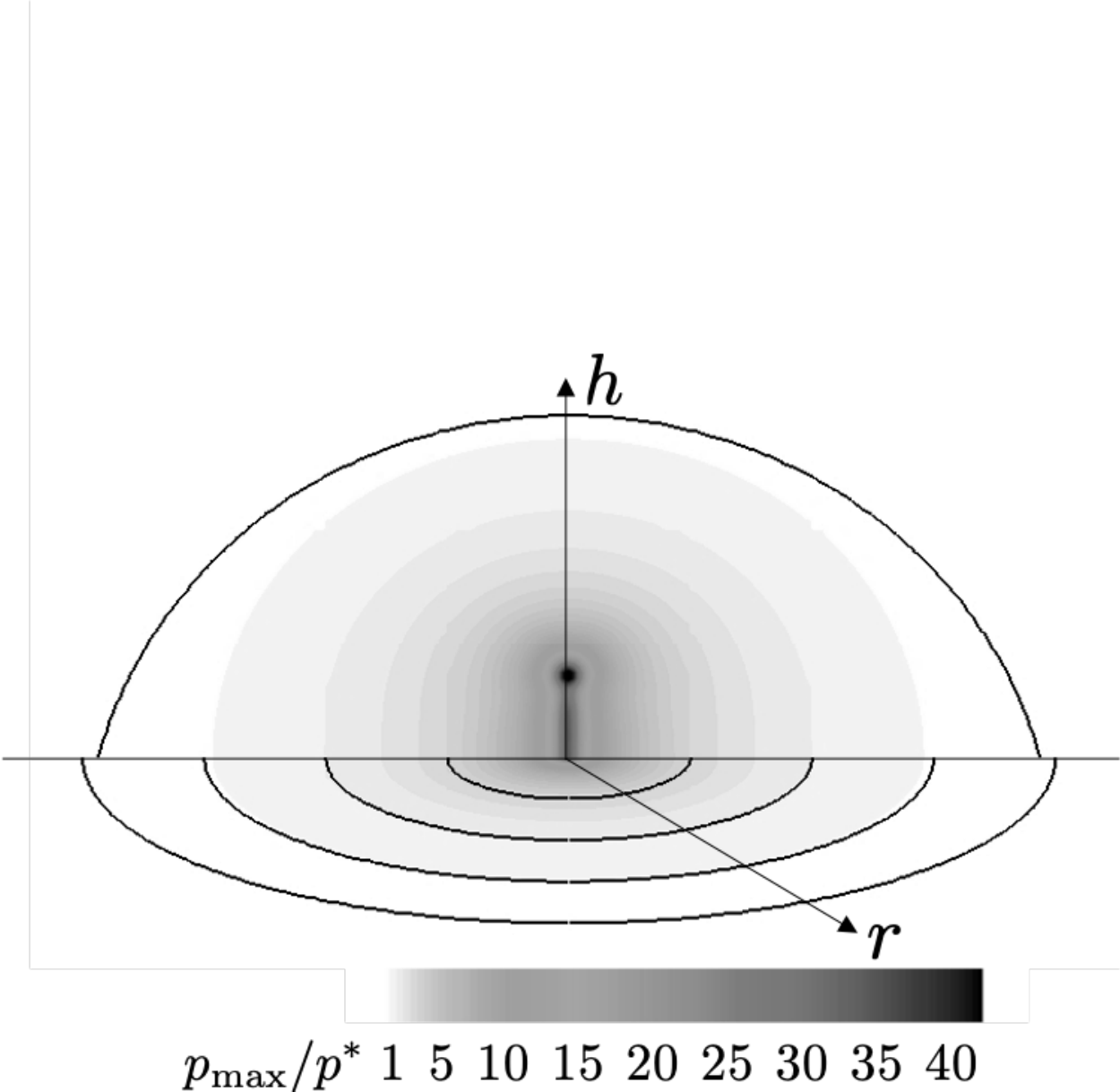}}
    \hspace{0.4cm}
    \subfigure[]{\includegraphics[height=6cm]{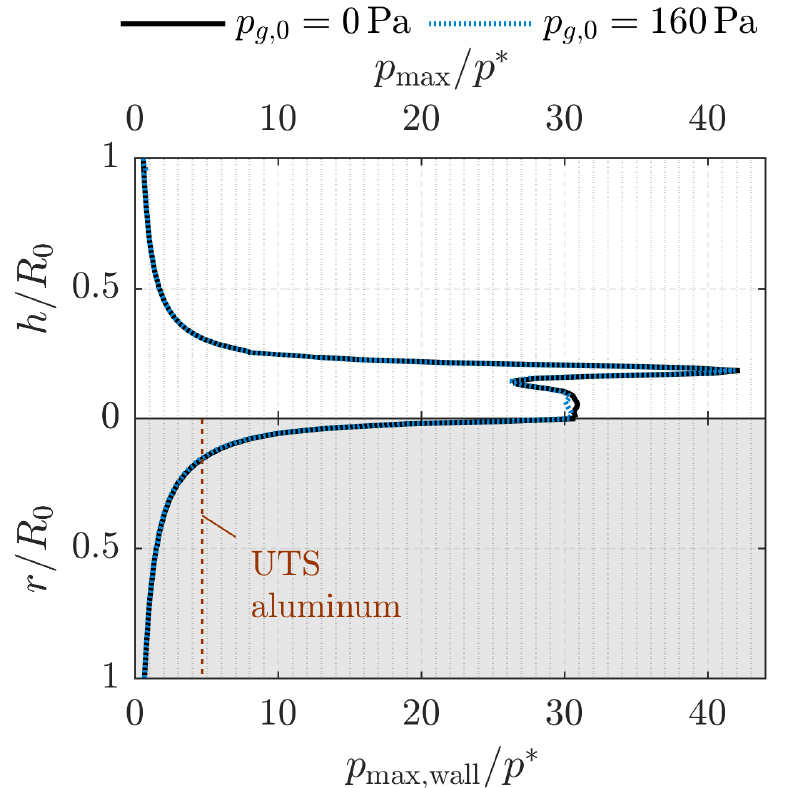}}  
    \caption{Maximum pressure induced by a collapsing bubble with $S/R_0=-0.25$.   (a) $p_\mathrm{max}/p^{\ast}$ on the wall (orientation rings at $r/R_0=0.25,0.5,0.75,1$) and midplane (initial bubble boundary indicated) (b) extracted $p_{max}/p^{\ast}$ and ultimate tensile strength (UTS) of aluminum ($70\,\si{\mega\pascal}$). }
    \label{fig:m0100_pmax}
\end{figure*}

\subsubsection{Wall-attached bubble with negative stand-off distance}
\label{sss:neg}

The collapse behavior of a vapor bubble with $S/R_0=-0.25$ is visualized in \cref{fig:m0100_ts}. Additionally, the comparison of a vapor bubble and a vapor-gas bubble for two selected time steps and a schematic representation of the collapse behavior are shown. The corresponding temporal evolution of the bubble volume and the recorded pressure signals in the wall-center are in the bottom of the figure. 

The wall-attached bubble is pinched circumferentially at its maximum expansion, resulting in a mushroom shape (\cref{fig:m0100_ts}~(ii)). Such behavior was also reported by \citet{Shima:1977df} and \citet{Lauer:2012jh}. Additionally, a circumferential pinching has been also observed for ellipsoidal bubbles \citep{Pishchalnikov:2018pp, Lechner:2019fp}. The radially inward directed flow reaches very high velocities, here exceeding $200\,\si{m/s}$ ($\approx 20 u^{\ast}$). Later, the collision of the waterfronts induces a high pressure peak, which can be seen in the pressure signals (\cref{fig:m0100_ts} bottom left). Shortly afterward the remaining upper part (the 'mushroom head') collapses, emitting a shock wave. When this wave reaches the wall, it induces another increase of the pressure signals (\cref{fig:m0100_ts} bottom left). Thus, the collision of the waterfronts and not the collapse is the central mechanism for the maximum wall pressure, which has also been observed for high driving pressures~\citep{Trummler:2021if}. Due to the conservation of momentum, the preceding radial inward flow at the pinching now causes a flow in upward direction reaching more than $100\,\si{m/s}$ ($\approx 10 u^{\ast}$), see \cref{fig:m0100_ts}~(iv). The rebound takes place in the shear layer resulting in a vapor torus (\cref{fig:m0100_ts}~(v)).

If gas is present in the vapor bubble, the collapse is slightly decelerated and a higher gas content occurs at the boundary where the vapor has already collapsed, see \cref{fig:m0100_ts}~(iii')(iv'). The gas decelerates the circumferential pinching and reduces its velocity by 3.25\%. The reduced velocity correlates with a damped maximum pressure at collision (see \cref{fig:m0100_ts} bottom left). As expected, the rebound with gas is stronger, as visualized in \cref{fig:m0100_ts}~(iv'). 

\Cref{fig:m0100_pmax} shows the distribution of the maximum pressure on the mid-plane and the wall. The highest pressure occurs at the focus point of the collapse. The high pressure along the symmetry line and in the center of the wall is due to the collision of the liquid fronts. The gas dampens the maximum pressure at the focus point by 0.8\%, which corresponds to the damping effect at a spherical collapse (see \cref{s:spherical})\new{, and the maximum wall pressure by 1.34\%.}

Based on the maximum wall pressures, material damage can be estimated. In experiments aluminum with an ultimate tensile strength (UTS) of about $70\,\si{\mega\pascal}$ ($\approx 4.7p^{\ast}$)~\citep{malmberg2015aluminium} is often used. Taking the UTS as the threshold, the estimated wall damage for aluminum is indicated in \cref{fig:m0100_pmax}, and would be a central, pit-shaped surface deformation. 

\subsubsection{Wall-attached bubble with positive stand-off distance}
\label{sss:pos}

\begin{figure*}
   \centering
    {\includegraphics[width=0.8\linewidth]{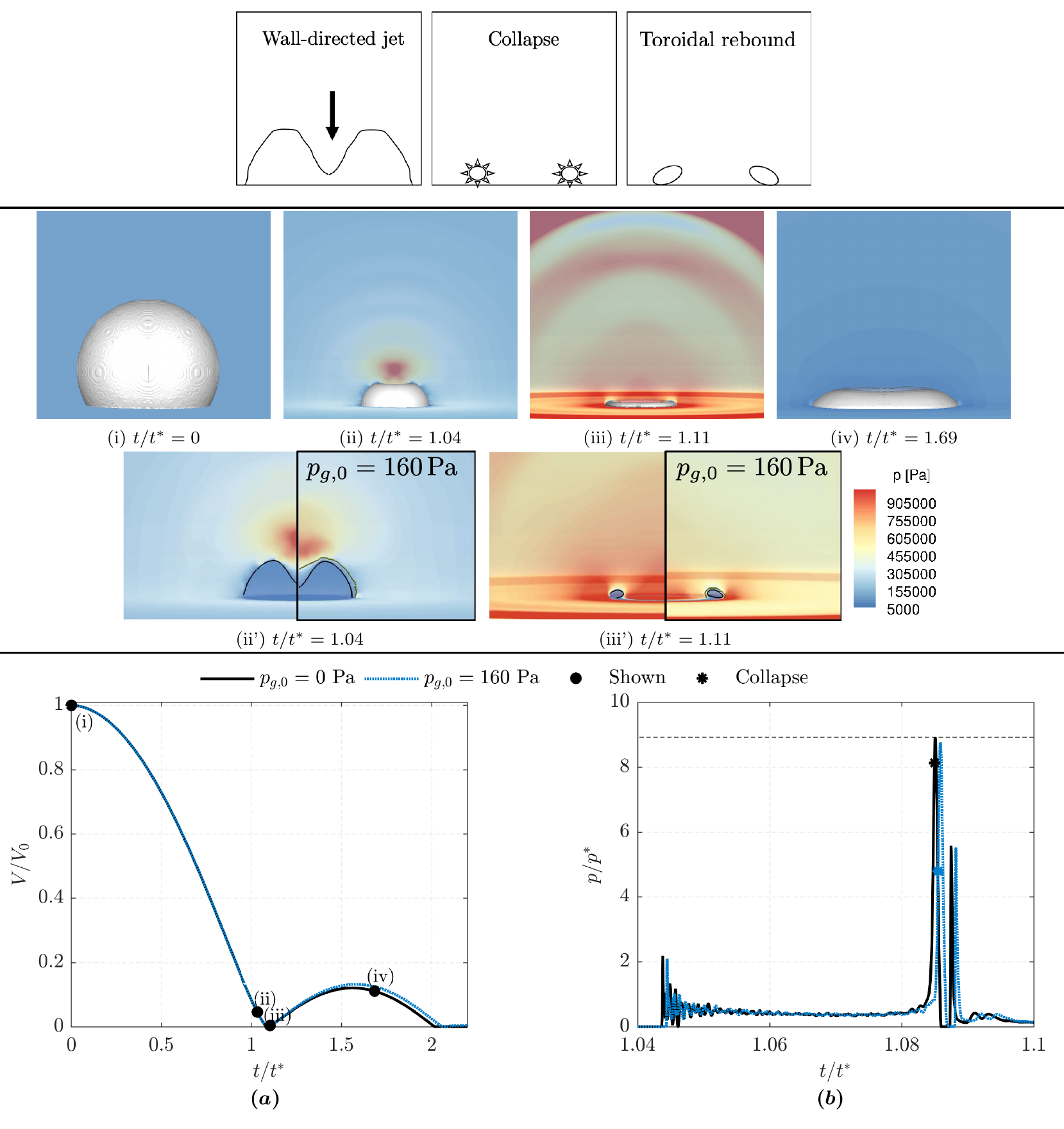}}
    \caption{Collapsing wall-attached bubble with $S/R_0=0.5$. Top \new{panel}: Sketch of general collapse behavior. Middle \new{panel}: Time series showing pressure on midplane and isosurface 10\% vapor [(i)-(iv)] and a comparison of $p_{g,0}=0\,\si{Pa}$ and $p_{g,0}=160\,\si{Pa}$ with additional isolines 10\% gas (olive) [(ii)-(iii)]. Bottom \new{panel}: Temporal evolution of bubble volume (left), and recorded pressure signals at the wall-center (right). \new{Reproduced from \citet{Trummler:2021diss}}. 
      }
      \label{fig:p0200_ts}
  \end{figure*}

  \Cref{fig:p0200_ts} visualizes the collapse of a vapor bubble. Additionally, the comparison of a vapor bubble and a vapor-gas bubble for two selected time steps, a schematic representation of the collapse behavior, the corresponding temporal evolution of the bubble volume and the recorded pressure signals in the wall-center are shown.

  In this configuration, the least resistance of the bubble is in the wall-normal direction and the surrounding pressure distribution leads to an indentation on the upper side. A wall-directed liquid jet forms, penetrating the bubble and resulting in a torus. Then the first collapse takes place, followed by a toroidal rebound and a second collapse. This behavior is well known and has been analyzed in several experimental~\citep{Tomita:1986gy,Philipp:1998eg} and numerical studies~\citep{Lauer:2012jh,Trummler:2021if}.

  The wall-centered pressure signals (\cref{fig:p0200_ts} bottom left) show the impact of the jet, followed by two pressure peaks induced by the first collapse. These peaks are significantly higher than the jet-induced one, which agrees with the literature~\citep{Lauer:2012jh,Philipp:1998eg}. 

  The presence of gas, again, results in a higher gas content at the boundary, see \cref{fig:p0200_ts}~(ii'),(iii'). Furthermore, it delays the first collapse, leads to a stronger rebound and a delayed second collapse (\cref{fig:p0200_ts}~bottom). The gas attenuates the velocity of the wall-directed jet and thus the intensity of the jet impact by 3.91\%, as can be seen in the pressure signals. The collapse induced pressure peak is also damped by the gas.

  \begin{figure*}
      \centering
    \footnotesize
      \subfigure[]{\includegraphics[height=6cm]{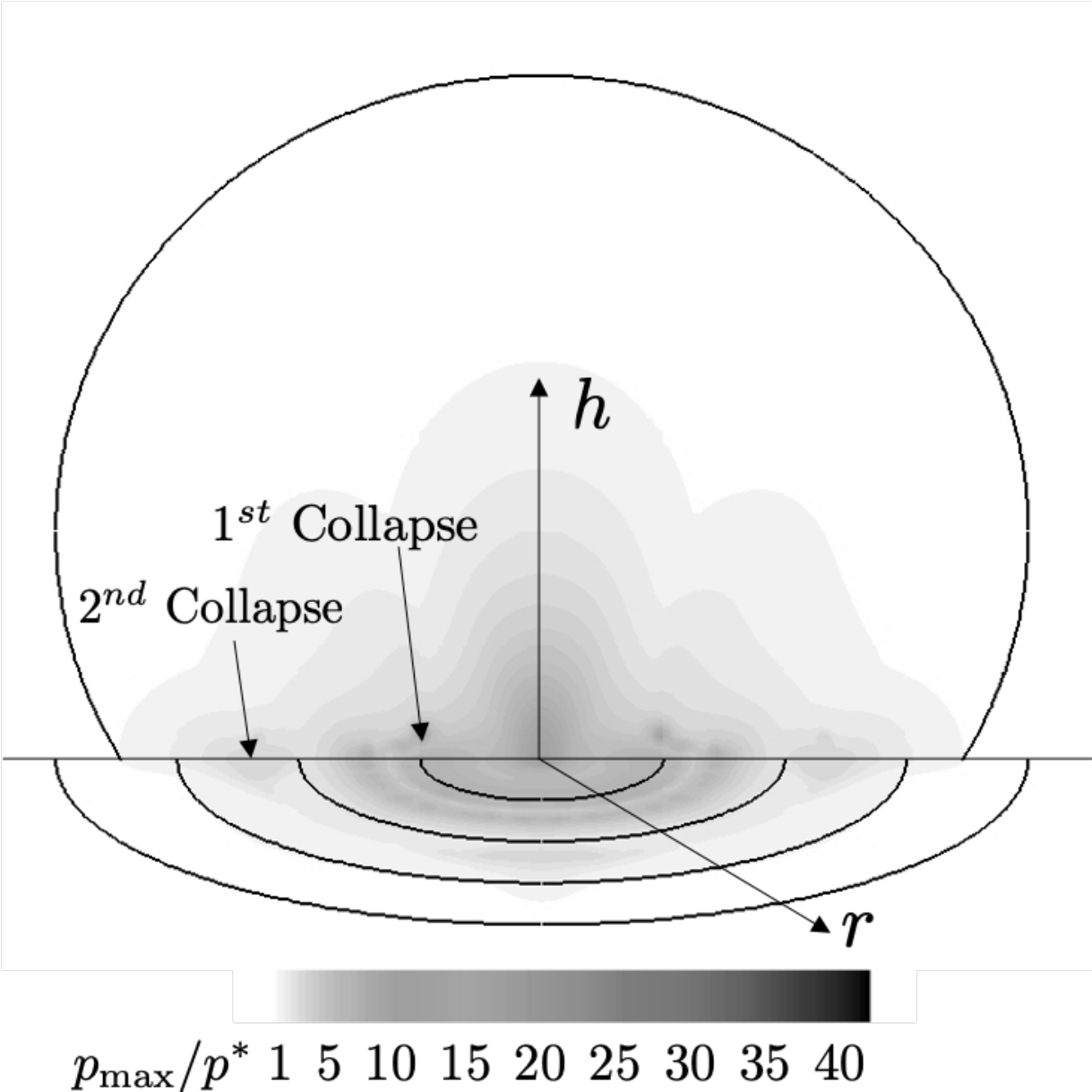}}
      \hspace{0.4cm}
      \subfigure[]{\includegraphics[height=6cm]{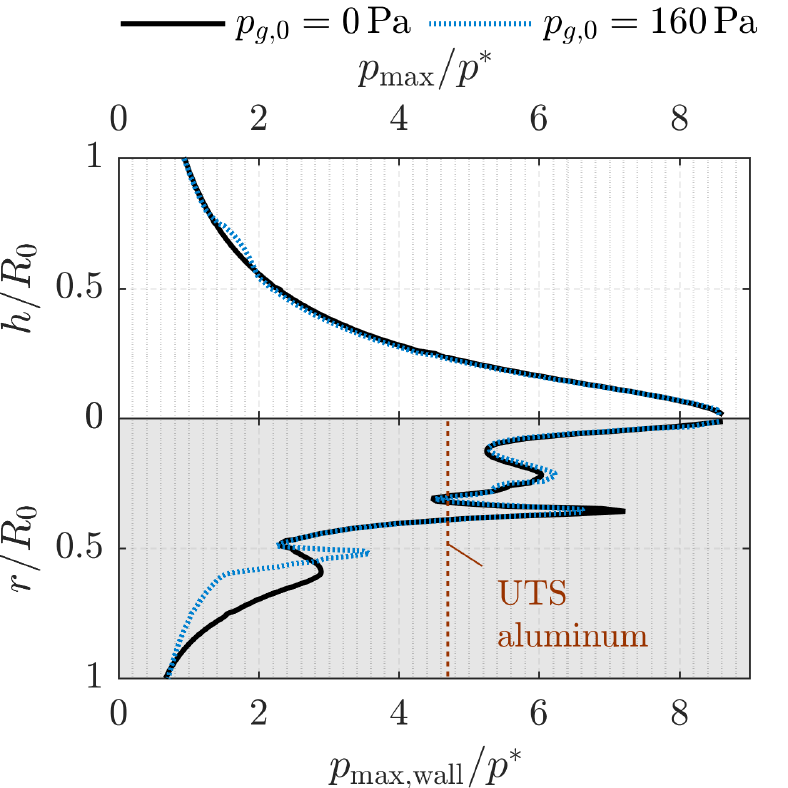}} 
   \caption{Maximum pressure induced by a collapsing bubble with $S/R_0=0.5$. (a) $p_\mathrm{max}/p^{\ast}$ on the wall (orientation rings at $r/R_0=0.25,0.5,0.75,1$) and midplane (initial bubble boundary indicated) (b) extracted $p_{max}/p^{\ast}$ and ultimate tensile strength (UTS) of aluminum ($70\,\si{\mega\pascal}$). }
  \label{fig:p0200_pmax}
  \end{figure*}

  The distribution of the maximum pressure is shown in \cref{fig:p0200_pmax}. The first collapse induces pressure peaks in the center of the collapsing torus and high pressures at the wall below. However, the highest wall pressure is recorded in the center and induced by the superposition of the emitted pressure waves at first collapse. The second collapse takes place radially further outwards and causes significantly lower wall pressures than the first collapse. If gas is present, the maximum wall pressure induced by the second collapse is higher and shifted radially inwards. Otherwise, the gas slightly dampens the maximum pressures. The damping of the maximum wall pressure at first collapse is 6.6\%, which is significantly higher than in the other configurations. On an aluminum specimen, the collapse would probably lead to ring-shaped damage with radius $r=0.35R_0$ and an indentation in the center, which matches experimental observations~\citep{Philipp:1998eg}. 

  The maximum wall pressure at $S/R_0=0.5$ is about a third of the one at $S/R_0=-0.25$. This is consistent with the observations of \citet{Lauer:2012jh,Trummler:2021if}, who found that the maximum wall pressure decreases with increasing stand-off distance and that this decrease is less pronounced for negative distances. 

\section{Conclusions and Discussion}
\label{s:ConclusionAndDiscussion}

  We have suggested a modified multi-component model to simulate vapor bubbles containing free, non-condensable gas. By numerical simulations, we were able to reproduce the physical effects of gas inside a vapor bubble. Free gas in a vapor bubble leads to a stronger rebound and dampens the emitted shock wave. This effect is already visible with coarse grid resolutions but is more pronounced for higher grid resolutions. Additionally, we were able to reproduce the partitioning into rebound and shock wave energy proposed by~\citet{Tinguely:2012wo} and could confirm a $\psi$-equivalence. This validation enabled us to investigate the effect of free gas inside vapor structures on more complex configurations such as the collapse of wall-attached bubbles. 

  The second part of the paper presented simulation results of a collapsing wall-attached bubble under atmospheric pressure. We investigated the collapse behavior and pressure impact for the selected stand-off distances $S/R_0=-0.25$ and $S/R_0 = 0.5$. The observed collapse behavior resembles that of previous investigations at higher driving pressure differences. Our simulation results provide deeper and additional insights into the rebound behavior and the relevant mechanisms for pressure peaks. We showed that at $S/R_0=-0.25$ the collision of the circumferential pinching induces the maximum wall pressure and not the final collapse. At $S/R_0 = 0.5$, we captured the first and second toroidal collapse and the induced wall pressures of both. The induced wall pressure of the second collapse is weaker and radially further outward. 

  For aspherical collapses \new{under atmospheric conditions, we observed a small effect of the non-condensable gas in our simulation results. Direct comparison of a collapsing vapor-gas bubble with a collapsing vapor bubble showed that the presence of gas slightly decelerates the collapse and reduces the velocity of the liquid jets, i.e. the circumferential pinching or the wall-directed jet. As expected,} gas dampens the collapse pressure and enhances the rebound. We found that the damping of the maximum wall pressure by the gas depends on the mechanism inducing this pressure peak. \new{ In case of a toroidal collapse, the observed damping of the maximum wall pressure is 6.6 \%, while for the bubble with a negative stand-off distance it is 1.35 \%. }

  Nevertheless, our findings for the gas effect in aspherical configurations might be biased by the employed isothermal modeling of the gas. Since we model the gas isothermal, we initialize a relatively high gas content of $p_{g,0}=160\,\si{Pa}$ compared to the assumed one of $p_{g,0}=3-10\,\si{Pa}$. The $\psi$-equivalence, which justifies this initial value, was only shown for spherical collapse. To evaluate the effect of the gas in detail, adiabatic modeling of the gas has to be employed. 
  Moreover, further experimental and numerical investigations are generally necessary to quantify the effect of gas inside vapor bubbles. A major uncertainty of these studies is that the actual gas content in practical applications is generally unknown and very difficult to estimate. 

  \new{The numerical framework presented can be used to study the effects of gas in configurations of interest. Accurate knowledge of the gas effect in aspherical collapses allows precise control of the effects on collapse pressure, or respectively destruction potential, and rebound behavior. Such knowledge can be advantageous for e.g. biomedical applications, such as urinary stone ablation~\citep{Pishchalnikov:2018pp}, needle-free injection with pressurized auto-injectors~\citep{Veilleux:2018gy}, or new technologies, such as surface-cleaning~\citep{Reuter:2017bu} and micro pumps driven by bubble rebound~\citep{dijkink2008laser}. Furthermore, the findings can also be applied to control erosion aggressiveness~\citep{Schmidt:2014ev}. } 

\section*{Acknowledgment}

  The authors gratefully acknowledge the Gauss Centre for Supercomputing e.V. (www.gauss-centre.eu) for funding this project by providing computing time on the GCS Supercomputers SuperMUC and SuperMUC-NG at Leibniz Supercomputing Centre (www.lrz.de).

\section*{Data Availability}

  The data that support the findings of this study are available from the corresponding author upon reasonable request.

  \bibliography{pof}
  \bibliographystyle{elsarticle-harv}
  \biboptions{authoryear}

\end{document}